\documentclass[preprint,12pt]{aastex}

\usepackage{color}
\usepackage{natbib}

\newcommand{\onvire}[1]{}

\begin{document}
 
\title{The effects of discreteness of galactic cosmic rays sources}

\author{R. Taillet\altaffilmark{1,2}}

\author{P. Salati\altaffilmark{1,2}, D. Maurin\altaffilmark{3,4}}                
\author{E. Vangioni-Flam\altaffilmark{3} \and M. Cass{\'e}\altaffilmark{3,4}}
      
\affil{$^1$ Laboratoire de Physique Th{\'e}orique LAPTH,
      Annecy--le--Vieux, 74941, France}
\affil{$^2$ Universit{\'e} de Savoie, Chamb{\'e}ry, 73011, France}
\affil{$^3$ Institut d'Astrophysique de Paris, 98 bis Bd
      Arago, 75014 Paris, France}      
\affil{$^4$ SAp, CEA, Orme des Merisiers, 91191 Gif-sur-Yvette, France}

\shortauthors{Taillet et al.}

\shorttitle{Diffusion from discrete sources}
                        
\date{Received \today; accepted}

\begin{abstract}

Most studies of GeV Galactic Cosmic Rays (GCR) nuclei assume
a steady state/continuous distribution for the sources of cosmic rays, 
but this distribution is actually discrete in time and in space.
The current progress in our understanding of cosmic ray
physics (acceleration, propagation), the required consistency in explaining
several GCRs manifestation (nuclei, $\gamma$,...) as well as the precision
of present and future space missions (e.g. INTEGRAL, AMS, AGILE, GLAST)
point towards the necessity to go beyond this approximation.
A steady state semi-analytical model that describes well many nuclei data
has been developed in the past years based on this
approximation, as well as others. 
We wish to extend it to a time dependent version, including discrete sources.
As a first step, the validity of several approximations of the model
we use are checked to validate the approach: i) the effect
of the radial variation of the interstellar gas density
is inspected and ii) the effect of a specific modeling for
the galactic wind (linear vs constant) is discussed. In a second step,
the approximation of using continuous sources in space is considered.
This is completed by a study of time discreteness through
the time-dependent version of the propagation equation.
A new analytical solution of this equation for instantaneous point-like sources,
including the effect of escape, galactic wind and spallation, is presented.
Application of time and space discretness to definite propagation conditions and
realistic distributions of sources will be presented in a future paper.
\end{abstract}

\maketitle


\section{Introduction}

The cosmic ray flux at any given position in the Galaxy is due to
many sources, which are probably 
related to the remnants of supernovae.
Each source of position $\vec{r}_i$ and age $t_i$ yields,
at position $\vec{r}_0$ and time $t=0$ (now),
a flux $N(\vec{r}_i, t_i, \vec{r}_0, t_0=0)$  that can be obtained by 
solving the diffusion equation with
the appropriate source term and boundary conditions (see below).
The total flux is given by
\begin{equation}
      N_{\rm tot}(\vec{r}_0)= \sum_i N(\vec{r}_i, t_i,\vec{r}_0)
      \;.
      \label{somme_stat}
\end{equation}
This model has been coined Myriad model by \citet{Higdon03}.

The effects of discreteness have been studied in the past.
As regards the spatial discreteness, \citet{Lezniak79} start with the time 
dependent diffusion equation when both spallations and energy losses are 
taken into account, to eventually derive the steady-state Green 
function necessary to study the no-near source effect (expected
to reproduce a depletion of the path length distribution at low grammage).
As for the temporal discreteness, \citet{Owens76} derived time dependent solutions in the framework of halo models (see also \citealt*{Lezniak79a}), 
including spallations but assuming a  gas density which is constant 
throughout the diffusive volume (in this case the mean time and the mean 
matter crossed are proportional). 
The most complete work to provide ``simple"
formula to the time-dependent case is probably 
\citet{Freedman80}, who derive the mean age and
the grammage distribution to seek if it is possible to constrain 
the propagation parameters from current observations of charged
nuclei. They conclude that a large degeneracy in propagation parameters 
remains in most cases.
It has been shown that the distribution of electrons and positrons at high energy 
is particularly sensitive to nearby sources, due to
their huge energy losses (\citealt{aharonian95}, see also
\citealt{Cowsik79}).
A nearby source such as the supernova remnant (SNR) associated with the Geminga
pulsars probably has a great influence on them.
The situation is less clear about the stable charged cosmic ray spectra,
but nearby sources are expected to be more and more important as energy
grows. Whereas some authors estimate the contribution from Geminga
to be at most 10\% \citep{johnson94}, some others \citep{Erlykin01a}
argue that a single supernova event can explain the feature
observed in the spectrum at a few PeV. This could be the
Sco-Cen association that is expected to be 
able to generate the Local Bubble 11 Myr ago \citep{benitez02}.
This can be tested by measuring the anisotropy
(see e.g. \citealt{Dorman85}), or by gathering observations from
the past and from the rest of the Galaxy. For instance, 
\citet{ramadurai93} argued that Geminga could be responsible
for an increase of the Cosmic ray flux by a factor 1.8 by inspection
of Antarctic ice sediments. The measurement of the isotopic
composition of the Earth crust (see e.g. \citealt{knie99}),
of meteorites, and of ice cores may be used to investigate
the time variation of the cosmic ray flux on Earth (see also
\citealt{Erlykin01b}). 

Most studies of the chemical composition of cosmic rays assume that 
the sum (\ref{somme_stat}) may be approximated by an integral
\begin{equation}
      N_{\rm tot}(\vec{r}_0) \approx \int d^2 \vec{r}
      \int_0^\infty dt N(\vec{r},t, \vec{r}_0)
      \; ,
      \label{integrale_stat}
\end{equation}
which is equivalent to (\ref{somme_stat}) only in the limit 
of a source distribution which is continuous
in space and time (continuous distributions for $\vec{r}_i$ and $t_i$).
This approximation is justified if the sources are numerous and 
densely distributed, but it probably fails for
nearby and/or recent sources, for which the detailed location and age 
should be known.
This paper is devoted to investigate the validity of this
approximation and to provide a more accurate description of diffusion when it
fails.
The chemical composition of cosmic rays is determined by the quantity of matter
that has been crossed by primaries during their propagation from the sources
to the Earth, and it can be conveniently described and studied by the 
grammage distribution.
Sec.~2 is devoted to the study of the grammage distribution in diffusion
models, and the importance of space discreteness of the sources on the
cosmic ray composition is investigated.
Sec.~3 presents a new analytical solution of the time-dependent diffusion
equation for point sources, taking into account spallations, convective wind
and escape. It is then used to find a criterion to separate the sources
in two categories, one containing the faraway and old, which can be modelled
by the usual steady-state model, the other containing the close or recent,
which require a finer description.
Sec.~4 concludes and presents some applications of the present work which will
be further developped in a next paper.

\section{Steady stade path-length distribution}

\label{sec:grammages}

During their journey between a source and Earth, Cosmic Rays cross 
regions in which interstellar matter is
present. 
The nuclear reactions (spallations) induced by the collisions lead to a change in the 
chemical composition.
Cosmic Rays can be sorted according to the {\em grammage}, i.e. 
the column density of matter they have crossed
(denoted by $x$ in this paper and usually expressed in g cm$^{-2}$).
If it is temporarily assumed that Cosmic Rays do not interact with 
the matter they cross, i.e. the spallations are switched off,
their density originating from a source located at
$\vec{r}_s$, detected at $\vec{r}_o$, and having crossed the grammage 
$x$ is called the {\em path-length distribution}.
It can be used to compute the probability of nuclear reaction when the spallations are switched on, and thus
it provides a tool to compare several diffusion models, as similar
path-length distributions give rise to similar chemical compositions.

\subsection{Definition - Generalized diffusion equation}

The evolution of the {\em path-length distribution}
${\cal G}(\vec{r}, t, x)$ at position $\vec{r}$ and time $t$ can be described by a generalized 
diffusion equation inspired by \citet{Jones79}:
\begin{equation}
     \frac{\partial {\cal G}(\vec{r}, t, x)}{\partial t} = K \Delta 
        {\cal G}(\vec{r}, t, x) + q(\vec{r}, t) \delta(x)
     - \frac{\partial {\cal G}(\vec{r}, t, x)}{\partial x}  m v n_{\rm 
        ism}(\vec{r})
     \; ,
     \label{eq_fond}
\end{equation}
where in the right-hand side, the first term stands for diffusion, 
the second is the source term (creating
particles with null grammage) and the third gives the augmentation of 
the grammage due to the crossing of matter.
It should be noted that when the density $n_{\rm ism}$ is not 
homogeneous in the whole diffusive volume,
grammage is not proportional to time (this will be discussed further 
in Sec.~\ref{subsec:higdon}).
We emphasize that Eq.~(\ref{eq_fond}) does not take into account the 
influence of the spallations on propagation,
but rather introduces the variable $x$ as a counter which keep tracks 
of the quantity of matter crossed by the Cosmic Rays, spallations being 
switched off.

The density can finally be obtained from the grammage distribution by noting
that a primary cosmic ray 
having crossed a grammage $x$ has a survival
probability given by $\exp(-\sigma x / m)$, where $\sigma$ is the
destruction cross-section.
In the following, $m$ will denote the {\it mean} mass of the 
interstellar medium atoms.
The probability for a Cosmic Ray emitted in $\vec{r}_s$ and reaching 
$\vec{r}_0$ unharmed is written as
\begin{equation}
     N_p(\vec{r}_s, \vec{r}_o) = \int_0^\infty \exp\left\{ - 
        \frac{\sigma}{m} x \right\}
     \, {\cal G}(\vec{r}_s, \vec{r}_o,x) \, dx
     \; .
     \label{laplace_transform}
\end{equation}
For a secondary species, a similar expression can be written
\begin{equation}
     N_s(\vec{r}_s, \vec{r}_o) = \int_0^\infty g_s(x)
     \, {\cal G}(\vec{r}_s, \vec{r}_o,x) \, dx
\end{equation}
where the function $g_s$ is obtained by solving the set of equations
\begin{equation}
        \frac{dg_s}{dx} = - \frac{\sigma_s}{m} g_s
        + \sum_i \frac{\sigma_{i \rightarrow s}}{m} g_i
\end{equation}
with the initial conditions (the values of $g_s$ for $x=0$) set to the source abundance of the 
considered species.

The path-length distribution, along with the B/C ratio, are computed in the next paragraph.
We take the opportunity of this computation to consider a more general situation than in our previous 
works, namely by considering a realistic radial dependence of the matter density in the disk.
The aim is twofolds : first, we want to comment on the computation of path lengths by \citet{Higdon03}, and in 
particular we want to discuss the existence of a feature at $x\sim 2$ g/cm$^2$ which they claim is present 
because of the H$_2$ ring in the galactic disk.
Second, this provides a way to effectively take into account the radial dependence by 
computing the mean matter density which is probed by each cosmic ray species, and to introduce this
mean density back in our code.

\subsection{Analytical result for radial distribution of matter}

\subsubsection{Path Length Distribution (PLD)}

We compute the grammage distribution in the diffusion model we used in 
previous studies \citep{Maurin01, usine_rad}. 
It exhibits cylindrical symmetry,
escape happens through the $z=\pm L$ and $R=20$ kpc boundaries, 
galactic wind is constant in the halo and matter is localised in a thin 
disk at $z=0$. 
We consider a radial dependence of the surface mass density $\Sigma(r)$
taking into account the radial distributions of HI, HII and H$_2$ 
\citep{ferriere98}.
For convenience, we normalize this quantity by the local mean surface density
$\Sigma^0_{\rm ism} \equiv 6.2 \times 10^{20}$ cm$^{-2}$, i.e. we 
introduce $f(r) \equiv \Sigma(r)/\Sigma^0_{\rm ism}$.
The models used in our previous studies considered only flat matter distribution, to keep the problem 
tractable in a semi-analytical way, and we take the opportunity of this study to investigate the importance
of this assumption.

The generalized diffusion equation~(\ref{eq_fond}), with the left-hand side set to 0,
can then be solved as detailed in Appendix~B, by expanding the quantities over 
a set of Bessel functions. The solution is
\begin{equation}
    {\cal G}(r,z, x) = \sum_{i=0}^\infty 
    \frac{\sinh \left(\zeta_i (L-|z|)/R
    \right)}{\sinh \left(\zeta_i L/R \right)} \times
    J_0 \left(\zeta_i \frac{r}{R} \right)
    \sum_{j=0}^\infty a_{ij} e^{-x/x_j} \, \Theta(x)\,
    \; ,
    \label{sol_non_hom}
\end{equation}
where $\Theta(x)$ is the Heaviside distribution, the $a_{ij}$ and the $x_j$ are the 
eigenvectors and eigenvalues of the matrix
\begin{displaymath}
     A_{ij} = \frac{2m v \Sigma^0_{\rm ism} }{KS_i J_1^2(\zeta_i)} 
     \tanh \left( \frac{S_i L}{2} \right)
     \int_0^1
     J_0(\zeta_i \rho) J_0(\zeta_j \rho) f(\rho r) \, d\rho
     \;\;\; \mbox{with} \;\;\;
     S_i = \frac{2\zeta_i}{R}
     \; ,
\end{displaymath}
For a flat distribution of matter ($f$ independant of $r$), this expression reduces to
\begin{displaymath}
     A_{ij} = \frac{m v \Sigma^0_{\rm ism} }{KS_i} 
     \tanh \left( \frac{S_i L}{2} \right) \, \delta_{ij}
\end{displaymath}
We have not considered energy losses in this computation, as the aim 
is not to provide a very sophisticated modelling of the cosmic rays 
diffusion, but rather to give an estimate of various effects.
{\color{blue} It follows that the results presented here do not apply 
directly to electrons and positrons, for which the energy losses are 
predominant.}

\subsubsection{Application}

\begin{figure}[h]
     \begin{center}
        \plotone{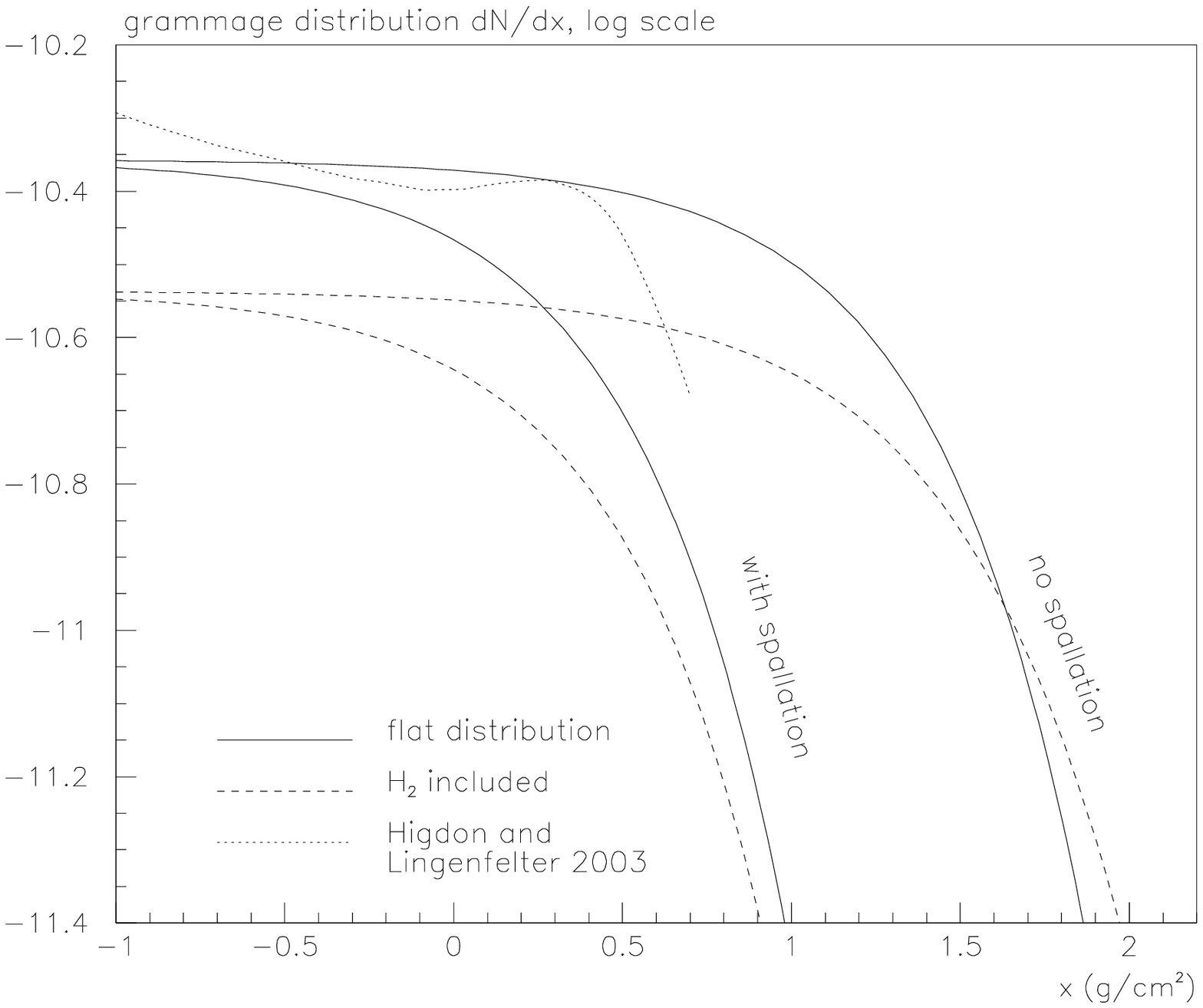}
        \caption{The upper curves represent the path-length distribution  ($x$ in g cm$^{-2}$) 
        for $L=4$ kpc and a homogeneous disk distribution corresponding to the local matter density, 
        and with a realistic gas distribution, including the
        H$_2$ ring, radially distributed according to \citet{ferriere98}.
        The lower curves are obtained by taking into account the spallation cross-section of Oxygen.
        The feature at $x\sim 2$ g cm$^{-2}$, visible in the Higdon points (dotted line) is not 
        reproduced by the analytical result.
        \label{influence_H2}}
     \end{center}
\end{figure}

We now use the above expression to evaluate the effect of the
choice of $\Sigma(r)$ on the path-length distribution, and then on the composition
of cosmic rays.
This is done by first computing the PLD for a flat and for a more realistic matter distribution.
The result for $K=0.03$ kpc$^2$/Myr and $L=5$ kpc is displayed in figure~(\ref{influence_H2}). 
The feature at $x\sim 2$ g cm$^{-2}$, visible in the Higdon points (crosses) is never reproduced 
by the analytical result. 
This difference is discussed in the next paragraph.

\begin{figure}[h]
     \begin{center}
        \plotone{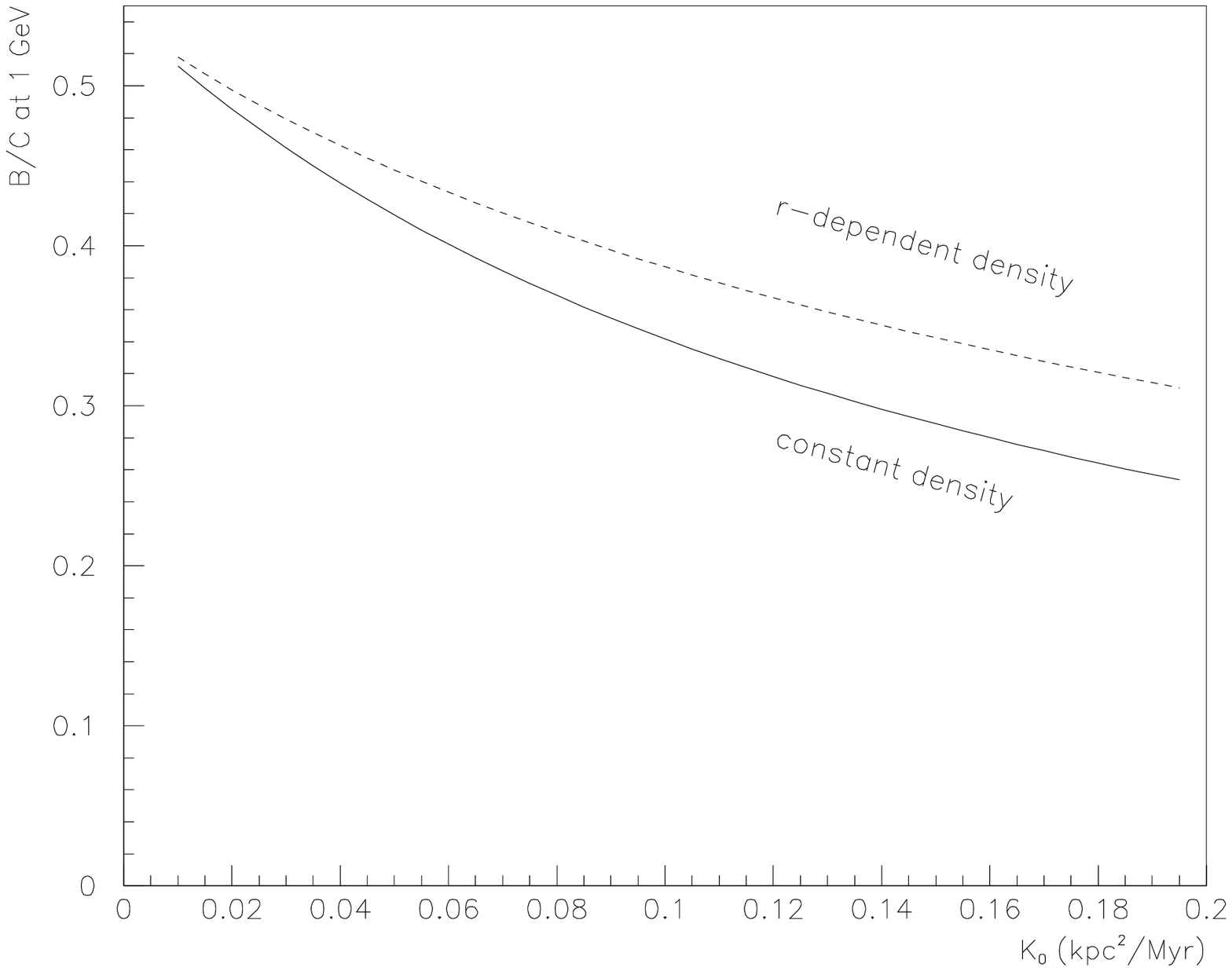}
        \caption{Secondary-to-primary ratio B/C, as a function of the diffusion coefficient normalisation
        $K_0$ in kpc$^2$/Myr, for a homogeneous matter distribution (solid line) and for
        a realistic gas distribution (dashed line).
        \label{influence_K}}
     \end{center}
\end{figure}

The effect on the composition of cosmic rays is illustrated in Fig.~\ref{influence_K}, 
where the B/C ratio  is computed in the two cases from the PLD.
For low values of the diffusion coefficient $K_0$, this ratio is not very sensitive to the global distribution
of matter, as the diffusion range is smaller, wheras for higher values of $K_0$, the B/C ratio is sensitive to
the increase of matter density at $r\sim 4$ kpc.
As the diffusion coefficient $K(E)$ actually increases with energy, the spectrum is likely to be affected by 
this effect, cosmic rays of higher energy probing a larger portion of the galactic disk.
However, for high values of $K$, the difference between a flat and a realistic distribution becomes 
independent of $K$, hence of $E$, so that the shape of the spectrum is not affected.
In particular, if $K_0$ is high, as hinted at by \citet{ptuskin_soutoul}, then the spectrum is not
affected by the presence of a H$_2$ ring.

These results provide an effective way to take into account the radial 
distribution of matter in our model. 
For a given energy, we can compute the constant interstellar gas density 
$n_H^{(B/C)}$ that must be assumed to reproduce the observed $B/C$ ratio.
The same thing can be done for sub-Fe/Fe, providing another 
effective density $n_H^{(Fe)}$. 
These densities are different because the corresponding species 
have different diffusion ranges, the latter being much more 
sensitive to spallations. The flux of each species can then be computed using 
the appropriate effective interstellar matter density.

\subsubsection{Comparison to the Higdon and Lingenfelter approach}
\label{subsec:higdon}

Another approach has recently been proposed in \citet{Higdon03} to compute the grammage
distribution: starting from the CR distribution $w(\vec{r},t)$ due 
to an instantaneous point source of age $t$, the mean matter density
seen at age $t$ by these CR is computed as
\begin{equation}
        n_{sn}(t) = \frac{\displaystyle \int \!\!\!\int \!\!\!\int d^3\vec{r}
        \, w(\vec{r},t) \, n_{ism}(\vec{r})}
        {\displaystyle \int \!\!\!\int \!\!\!\int d^3\vec{r}
        \, w(\vec{r},t)} \; .
        \label{average_higdon}
\end{equation}
Summing all the individual contribution with the appropriate weight, 
the mean matter density $n_{cr}(t)$ seen by the CR of age $t$ 
reaching the Earth is computed.
From this quantity, the evolution of the mean grammage is computed as
a function of time, through
\begin{equation}
        \frac{\partial \bar{X}}{\partial t} = \bar{m} \beta c n_{cr}(t)
        \label{eq4.2}
\end{equation}
which corresponds to their Eq.~(4.1) and (4.2).
We argue that this approach is not correct, for the following reasons.

First, the derivation of the time evolution of $\bar{X}(t)$ is given in Appendix~B,
and the equation (\ref{eq4.2}) is not recovered. The above expression would be
correct only with another (tricky) definition of $n_{cr}(t)$ and $n_{sn}(t)$.

Second, the averaging process (\ref{average_higdon}) gives the mean density
as seen by {\em all} the CRs emitted by the source, whereas what is needed would be the
mean density seen by the CRs that reach the Earth (those we do observe).
These quantities are different, and the corresponding time evolution is computed
in Appendix~B.

Finally, and most importantly, it is quite tricky to infer the grammage distribution
${\cal G}(\vec{r}, x)$ (needed to apply the weighted slab technique) 
from the mean grammage $\bar{X}(t)$, and 
their relation (4.3) is not correct. To see that more clearly,
consider the more fundamental quantity ${\cal G}(\vec{r}, x,t)$ giving,
at position $\vec{r}$, the 
density of CRs having crossed a grammage $x$ at age $t$.
The quantities introduced by \citet{Higdon03} are then related to 
${\cal G}(\vec{r}, x,t)$ through
\begin{equation}
        w(\vec{r},t) = \int_0^\infty dx\, {\cal G}(\vec{r}, x, t),
        \;\;\;
        {\cal G}(\vec{r}, x) = \int_0^\infty dt\, {\cal G}(\vec{r}, x, t),
        \;\;\;
        \bar{X}(t) = \frac{\displaystyle \int_0^\infty dx\, x\, {\cal G}(\vec{r}, x, t)}
        {\displaystyle \int_0^\infty dx\, {\cal G}(\vec{r}, x, t)}
        \label{redef_hidgon}
\end{equation}
which are fundamentally different from their (4.2) and (4.3). In particular, 
their $X(t)$ is actually $\bar{X}(t)$ and should be a function of 
$t$ in their (4.1) and (4.2), whereas the $X$ that appears in their grammage
distribution (4.3) should be a parameter and as such is {\em independent} on $t$.

This explains why we do not find the same grammage distributions as
\citet{Higdon03}. 
Their assimilation of the grammage distribution from individual sources 
to Dirac distributions has the effect of sharpening the final grammage distribution.
In particular, the feature at $x\sim 2$ g/cm$^2$ is not present in our results.
The distributions appear to be actually quite close to exponentials., 
i.e. to Leaky Box distributions.


\subsection{Spatial discreteness of the sources in a steady-state model}
\label{subsec:conclusion}

\subsubsection{General results}

We now want to investigate the effect of discreteness of the source
distribution on the Cosmic Ray composition, through the path-length
distribution. We first compute this quantity for a point source, and 
we then compare the path-length
distributions obtained for a set of point sources and an equivalent 
continuous source distribution.
For the sake of simplicity, we focus on the case of a uniform 
distribution of matter, for which $A_{ij}$ is diagonal and
the solution given in App.~B can be simplified.
As the composition of Cosmic Rays is only measured in the galactic
disk, we express the results in $z=0$.
It is then found that
\begin{displaymath}
        \tilde{\cal G}_i(z=0,x) = 
        \frac{q_i}{v \Sigma^0_{\rm ism}} \exp \left( -x/x_i\right)
        \Theta(x)
\end{displaymath}
with
\begin{displaymath}
     x_i = \frac{m v \Sigma^0_{\rm ism}}{2K} \left(\frac{1}{r_{\rm w}} 
+ \frac{S_i}{2} \coth \left( \frac{S_i L}{2}
     \right)\right)^{-1}
     \; .
\end{displaymath}
and $r_{\rm w} \equiv 2K/V_c$.
This expression could have been obtained by an inverse Laplace
transform of Fourier-Bessel coefficients of the steady-state density
(see e.g. \citealt{Maurin01})
\begin{displaymath}
     \tilde{N}_i(z=0) = \frac{q_i}{v \Sigma^0_{\rm ism} \sigma+ V_c +  KS_i \coth 
\left( S_i L/2 \right)}
     \; .
\end{displaymath}

The $q_i$ are obtained by Fourier-Bessel transforming the radial 
source distribution, which is assumed to be point-like and located in
the galactic disk. Unless this point source is at the galactic center, 
the cylindrical symmetry of the problem is broken and the previous
study does not apply.
However, as the influence of the $R$ boundary is expected to 
be negligible, we consider that the diffusion volume is not limited in 
the radial direction ($R \rightarrow \infty$).
The origin can then be set at the position of the source, which
restores cylindrical symmetry.
In this $R \rightarrow \infty$ limit,
the summations over Bessel functions become integrals,
the discrete sets $q_i$, $x_i$, $N_i$ become functions
$q(k)$, $x(k)$, $N(k)$ and the final 
result is obtained by performing the substitution
$1/J_1^2(\zeta_i) \rightarrow k\pi R/2$, $\sum_i \rightarrow  \int 
d(Rk/\pi)$ and $\zeta_i/R \rightarrow k$, 
\begin{equation}
     {\cal G}(r,z=0,x) = \frac{1}{v \Sigma^0_{\rm ism}} \int_0^\infty 
dk \, J_0(kr) \, q(k) \,
     \exp\left( - \frac{x}{x(k)}  \right)
     \label{grammage_source_ponctuelle}
\end{equation}
with
\begin{displaymath}
     x(k) = \frac{m v \Sigma^0_{\rm ism} /2K}{1/r_{\rm w} + S(k) \coth 
( S(k) L)}
\end{displaymath}
and
\begin{displaymath}
     S(k) = 2\left( \frac{1}{r_{\rm w}^2} + k^2 \right)^{1/2}
     \;\;\; , \;\;\;
     q(k) = k \int_0^\infty r\, dr \, \frac{\delta(r)}{2\pi r} J_0(kr) 
= \frac{k}{2\pi}
     \; .
\end{displaymath}
In the particular case of infinite $L$ and $r_{\rm w}= 0$, the expression
(\ref{grammage_source_ponctuelle}) gives
\begin{displaymath}
     {\cal G}(r,z=0,x) = \frac{1}{4\pi K r}
     \frac{K^2x}{m v^2 (\Sigma^0_{\rm ism})^2 r^2}\left( 1 + \frac{x^2 
K^2}{r^2 m^2v^2 (\Sigma^0_{\rm ism})^2} \right)^{-3/2}\, \Theta(x)
\end{displaymath}
which is the expression obtained in Taillet \& 
Maurin (2003) from a random walk approach.

\begin{figure}[h]
    \begin{center}
        \plotone{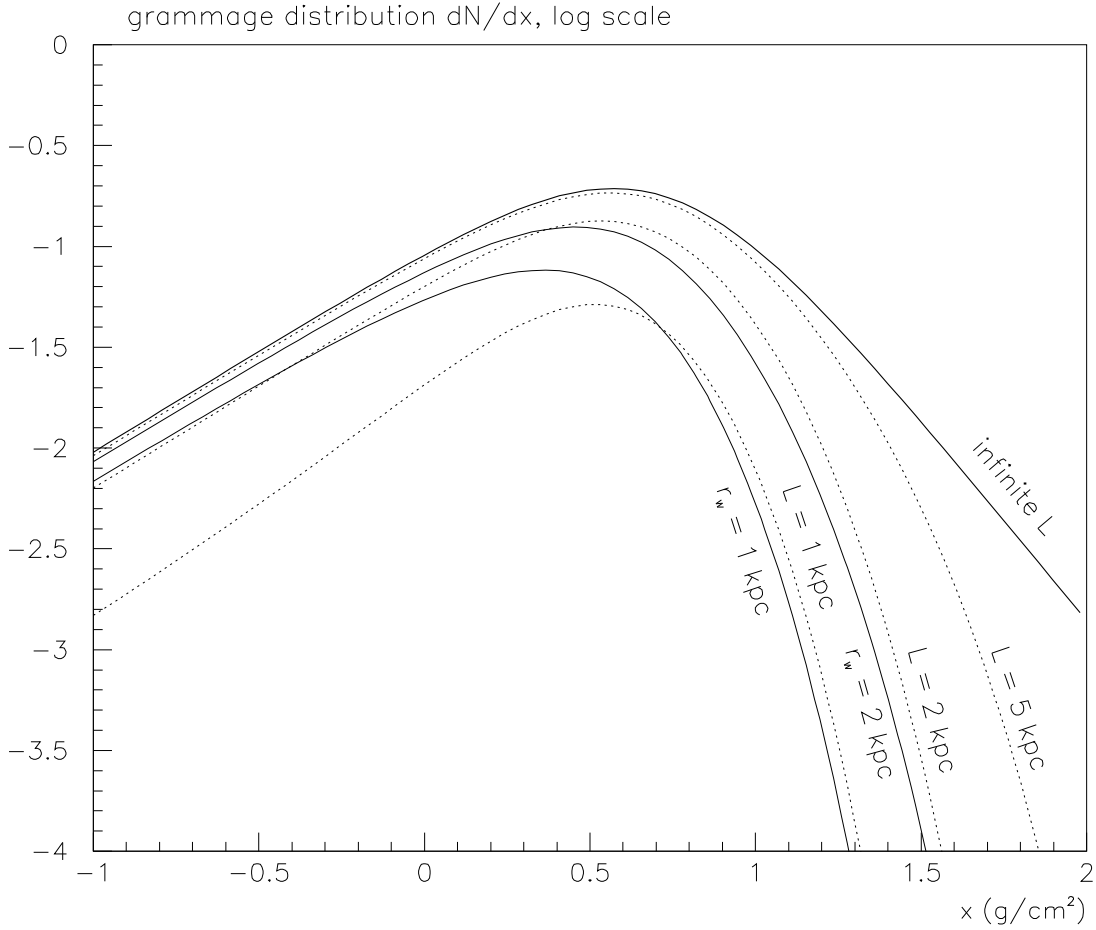}
        \caption{Path-length distributions ($x$ in g cm$^{-2}$) are shown for a point source
        located at a distance $r=1$ kpc, and for different propagation conditions.
        The thick line corresponds to free diffusion (no boundary, no wind).
        The thin dotted lines show the influence of escape for $L=1$, 2 and 5 kpc.
        The higher grammages, corresponding to longer paths, are more suppressed by escape, 
        which shifts the maximum of the path-length distribution towards low grammages.
        The effect of the convective wind, displayed as solid lines for $r_{\rm w}=1$ and 2 kpc, 
        is similar though quantitatively different for low grammages (see text for a discussion).
        The overall normalisation is decreased when escape and/or wind are considered, which reflects 
        the corresponding decrease of flux.
        \label{grammages_point_source1}}
    \end{center}
\end{figure}
\begin{figure}[h]
    \begin{center}
        \plotone{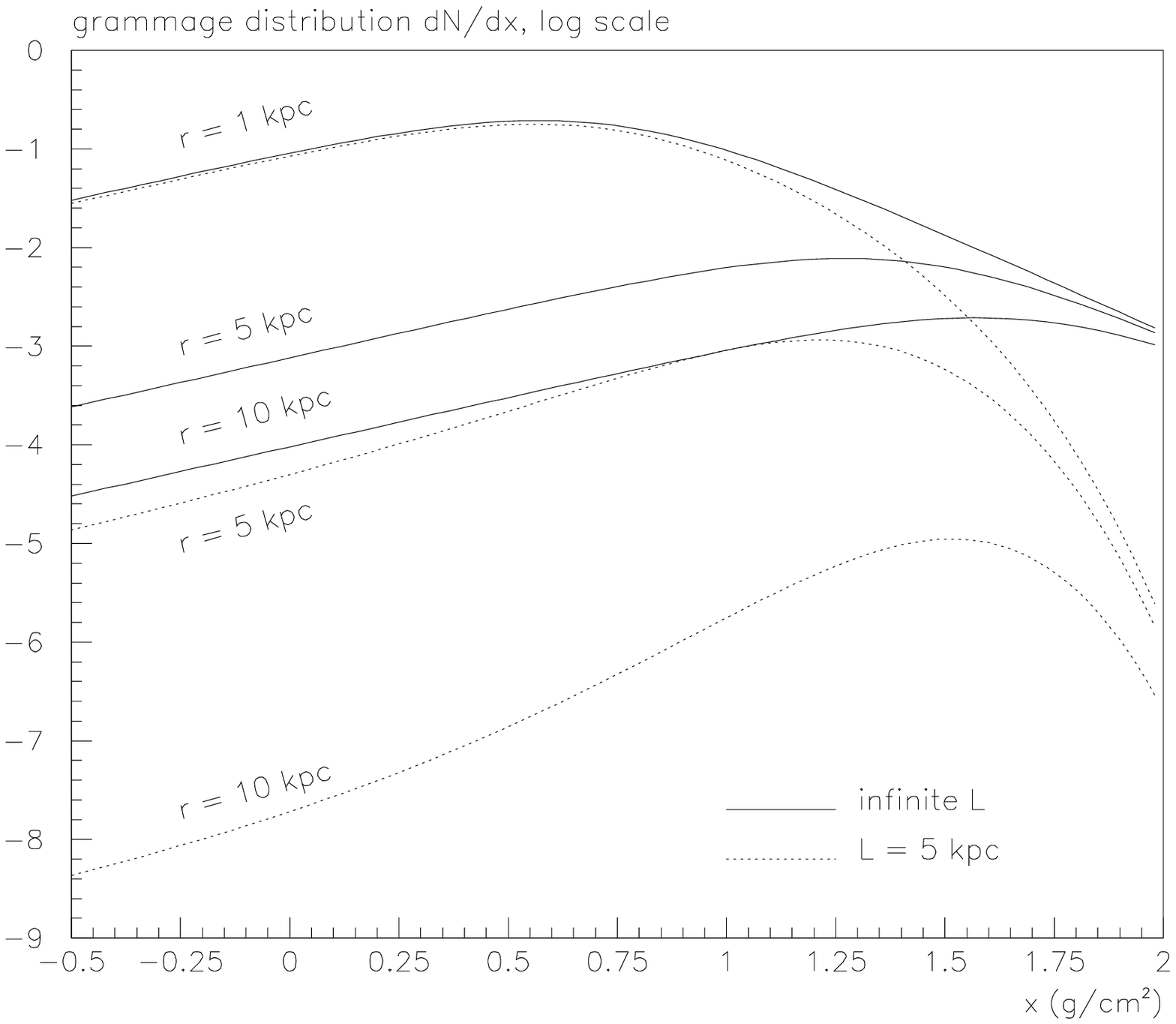}
        \caption{The path-length distribution ($x$ in g cm$^{-2}$) is 
        shown for a point source and for different $r$. The two situations $L=5$ kpc and $L$ infinite are shown for 
        comparison. The short grammages are depleted for remote sources, as the probability to travel a long 
distance 
        without crossing the disc is small. This effect is all more pronounced that $L$ is small.
        The normalisation also decreases with $L$, which reflects the corresponding decrease of flux.
        \label{grammages_point_source2}}
    \end{center}
\end{figure}

\subsubsection{Impact on the chemical composition}

The path-length distributions (\ref{grammage_source_ponctuelle}) are
displayed in Figs.~\ref{grammages_point_source1} and
\ref{grammages_point_source2}, for several propagation conditions, to
emphasize the relative effect of escape, galactic wind and spallation.
The effect of escape or wind is more important at 
higher grammages.
This was expected as they correspond to longer paths.
The effect of the convective wind is seen to be similar to that of
escape, but quantitatively different at low grammages.
To understand this, let us first consider diffusion in free space without wind. 
Several kinds of paths are responsible for low grammages: short paths, 
connecting us to nearby sources in the disk, and longer paths that
wander in the halo without crossing the disk too much. 
It happens that the second kind is rather important, which explains
why escape from a boundary at $z=\pm L$ (which kills the paths wandering 
too far in the halo) actually affects the low end of the grammage distribution. 
When wind is present, the short paths are more important, and the low
grammages are less affected.
        
Even for a point source, the path-lengths are quite broadly
distributed around the mean value, as can be
seen in Figs.~\ref{grammages_point_source1} and \ref{grammages_point_source2}.
As a result, the grammage distribution due to a set 
of discrete sources is smoothed to a great extent, and
that it is very unlikely to have observable consequences on the
composition of Cosmic Rays on Earth, except for very nearby sources (see below).
To illustrate this point, we compare the grammage distributions and the B/C ratio
from a smooth distribution to that obtained from a discrete sample representative 
of this distribution.
The relative importance of the sources located at different distances 
in making the observed composition (e.g. the B/C ratio) was estimated and discussed in
\citet{Taillet03}, and it was found that the nearby sources can be responsible for a 
substantial fraction of the flux of each species. 
The contribution of a point source located at a distance $r$ to the B and C flux, obtained
from (\ref{grammage_source_ponctuelle}), is displayed
as a function of $r$ in Fig.~\ref{grammage_source_ponctuelle}.
More specifically, we divide the sources into the far ($r>L$), the intermediate ($0.1L<r<L$) and
the nearby ($r<0.1L$), and Table~\ref{table_fraction} gives the fraction of flux coming from these regions, 
for different species.
\begin{table*}[ht]
  \begin{center}
    \begin{tabular}{|c|cccc|c|}   \hline
      &  B & C & sub-Fe & Fe & $n_{B/C}$(kpc$^{-2}$)\\
      \hline 
      $0<r/L<0.1$ & 3 \% & 12 \% &  7 \% & 20 \% & -- \\
      $0.1<r/L<1$ & 57 \% & 65 \% &  71 \% & 68 \%& 100  \\
      $1<r/L$ & 40 \% & 23 \% & 22 \% & 11 \%& 20 \\
      \hline
    \end{tabular}
    \caption{The first four columns give, for the four species
      indicated (B, C, sub-Fe and Fe), the fraction of flux 
      due to the three rings defined in the left column
      ($0<r/L<0.1$, $0.1<r/L<1$ and $1<r/L$), in the case of
      a homogeneous source distribution.
      The last column gives $n_{B/C}$, the surface number density of discrete 
      sources (in kpc$^{-2}$), randomly distributed in these rings, beyond 
      which the difference in flux with the continuous case is smaller 
      than 1 \%. 
      There is no number given for the inner ring because there is no 
      motivation to randomly draw the sources located very close to us,
      as these should be observed.
      The quantity $n_{B/C}$ was computed for $L=1$ kpc and should
      be divided by $L^2$ for other values. 
      It does not depend on the value of $K$.
      The effect of the convective wind can be estimated by 
      replacing $L$ by $L^\star \equiv K/V_c = r_{\rm w}/2$.}
    \label{table_fraction}
  \end{center}
\end{table*}
The grammage distribution from a ring delimited by $R_{\rm min}$ and $R_{\rm max}$
is given by
\begin{eqnarray}
        {\cal G}(r,z=0,x) &=& 
        \frac{1}{2\pi v \Sigma^0_{\rm ism}} \frac{2}{R_{\rm max}^2 - R_{\rm min}^2}\\
        &\times&
        \int_0^\infty dk \, \left\{ R_{\rm max}J_1(kR_{\rm max}) - R_{\rm 
        min}J_1(kR_{\rm min})\right\}\,
        \exp\left( - \frac{x}{x(k)}  \right) \; .
\end{eqnarray}
We randomly draw the position of sources inside each ring, with a 
given surface density of sources, we compute the corresponding flux, 
and the relative difference with the smooth case is computed.
As expected, this difference gets smaller as the source density 
is increased, i.e. as the granularity of the source distribution 
is decreased.
Table~\ref{table_fraction} gives $n_{B/C}$, the surface number density 
of sources in the disk (in kpc$^{-2}$) beyond which the difference in 
composition between the discrete and continuous case is less than 1 \%. 
The secondaries are less sensitive to the discreteness of sources, 
as their sources (the primaries) do have a continuous distribution.
These results do not depend on the value of $K$, as it is only 
sensitive to the relative importance of nearby and remote sources.

The results are presented as a function of $r/L$. 
If we now consider the disk defined by $r<1$ kpc, the resulting effect 
of granularity is quite insensitive to the value of $L$ (or $V_c$), 
as long as $L>1$ kpc (or $r_w >1$ kpc). 
This is because for $r<L$ (or $r<r_w$), the cut-off effect of 
escape (or wind) is always small.
As expected, the effect of discreteness is smaller for the outer rings.
The main effect is that when discrete sources are considered, the 
very nearby sources necessary to flatten the low end of the
distribution are always lacking. 
Another way of phrasing this result is that shot noise is 
dominated by the nearby sources.
The absence of nearby sources had been proposed by \citet{Lezniak79} 
to explain, in a different context, the depletion at low grammages 
that was thought to be observed, before \citet{Webber98} proposed a 
settlement to this particular issue.
The total path-length distribution is very close to an exponential, as
expected for a homogeneous source distribution in an infinite disk
\citep{Jones79}. 
\begin{figure}[h]
     \begin{center}
        \plotone{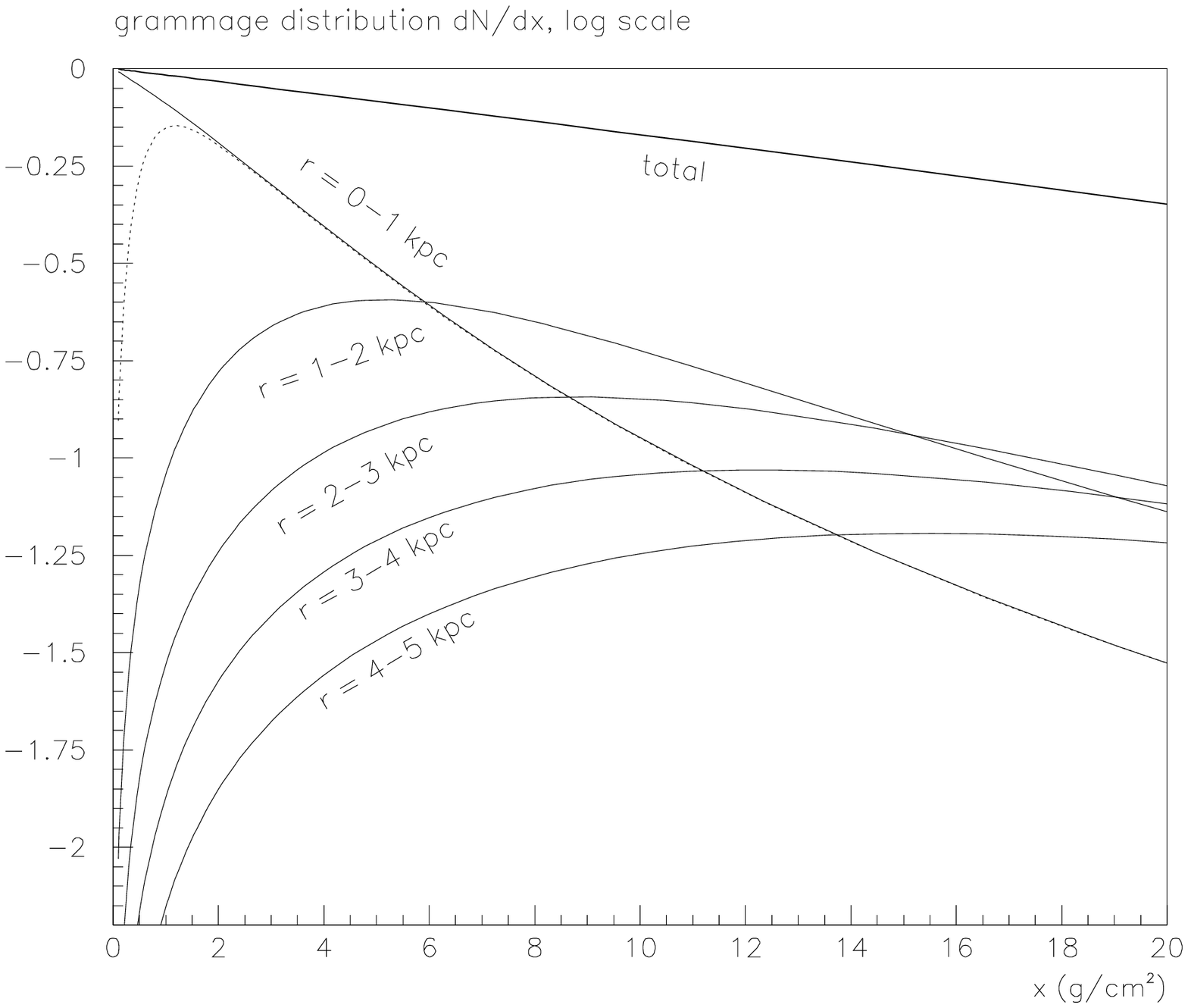}
        \caption{
        The contributions to the path-length distribution (thick line) 
	from continuous sources distributed in several rings (thin lines) 
	are displayed for $L=5$ kpc. The path-length distribution from a 
	set of 100 point sources drawn randomly in the central ring 
	(inner 1 kpc) is also shown (dotted line). The effect of 
        discreteness is most visible for low grammages. 
        For the other rings, the difference between the continuous 
        distribution and 200 discrete sources is of the order of a percent.
        The effect of galactic wind is very similar to that of $L$, replacing 
        $L$ by $L^\star \equiv K/V_c = r_{\rm w}/2$.
        \label{anneaux1}}
     \end{center}
\end{figure}
The effect of the wind is very similar to that of escape. It has been
remarked by \citet{Jones78} that to a given
value of $V_c$ (in the case where the wind velocity is constant in each side of the halo)
can be associated an effective escape height $L^\star = K/V_c = r_{\rm w}/2$, and we have
explicitly checked that the previous results apply by replacing $L$ by $L^\star$.


\subsection{Constant versus linear galactic wind}
\label{subsec:lin}

\begin{figure}[h]
     \begin{center}
        \plotone{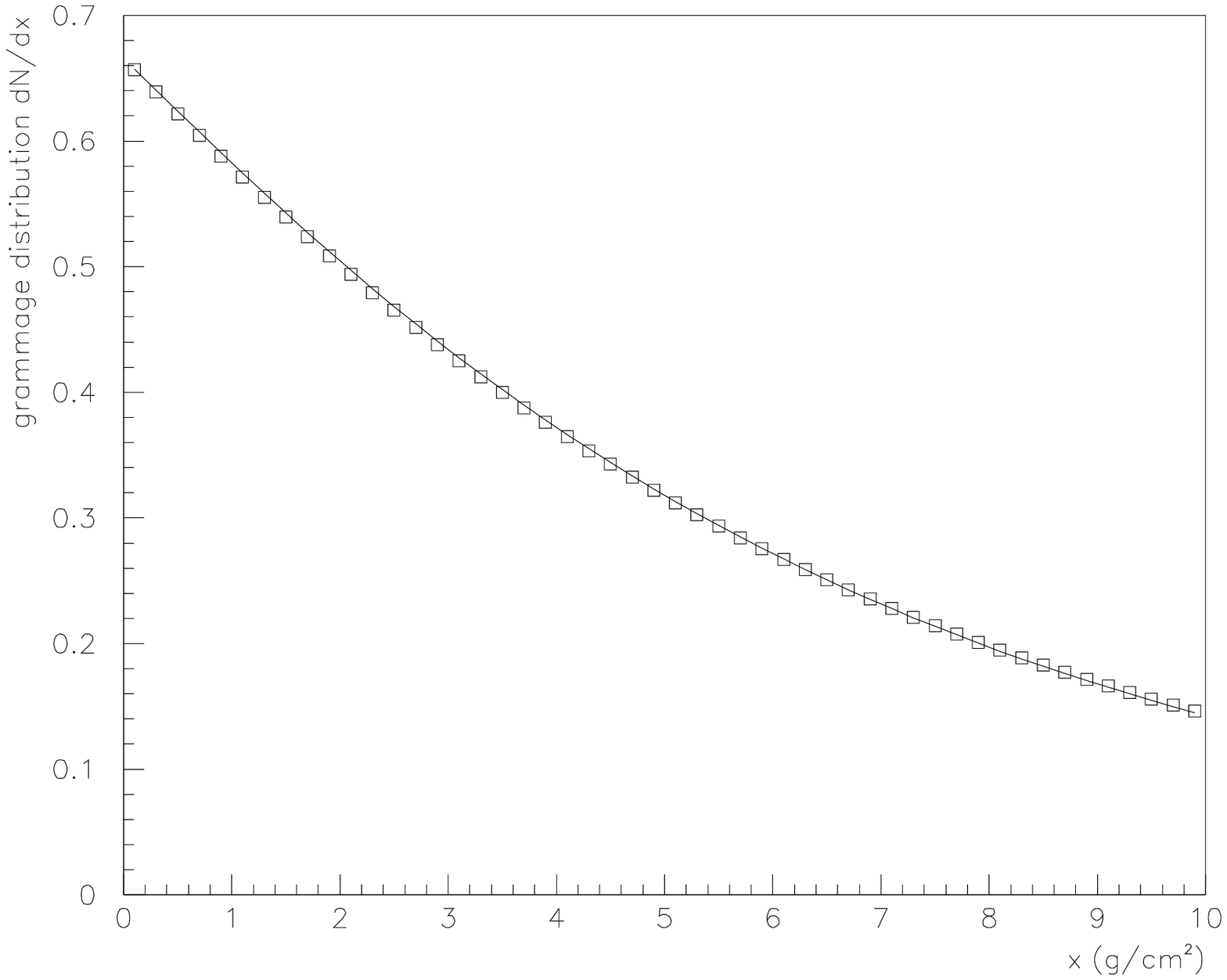}
        \caption{Path-length distributions for a linear (symbols, 
          $V_0 = 0.018$ kpc Myr$^{-1}$ kpc$^{-1}$)
        and a constant galactic wind (line, $V_c = 0.01$ kpc Myr$^{-1}$).
        It appears that the effect of a linear or constant wind 
        are very similar.
        For each value of $V_0$, it is possible to find a value of 
        $V_c$ yielding a grammage distribution which is
        very similar, with an accuracy better that one 
        percent.\label{grammages_vent_lin1}}
     \end{center}
\end{figure}

\begin{figure}[h]
     \begin{center}
        \plotone{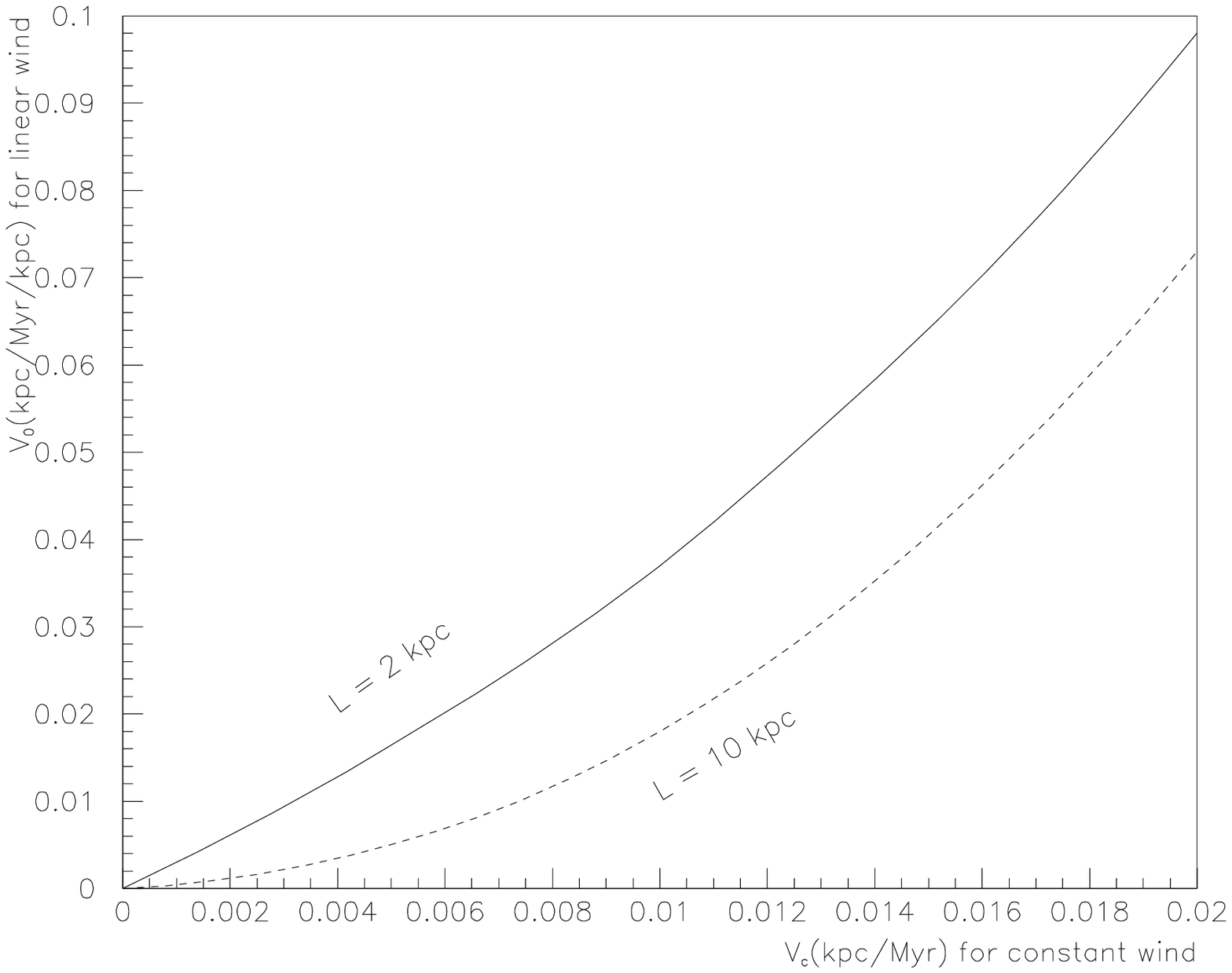}
        \caption{Correspondence between the values of $V_c$ 
	  for a constant wind ($x$ axis) and $V_0$ for a linear wind ($y$ axis)
          giving approximately the same path-length distributions, for 
	  two values of $L$.\label{grammages_vent_lin2}}
     \end{center}
\end{figure}

The previous results, as well as the results presented in our previous works,
rely on the assumption of a constant wind presenting a discontinuity
through the galactic disk.
To probe the sensitivity of our results to this hypothesis, we 
consider another model in which the
value of $V_c=V_0 z$ varies linearly with $z$. This corresponds to
the choice made e.g. in the widely used code {\sc galprop}.
The calculations are detailed in Appendix~C, and the
results will be used here without further justification.
The path-length distribution reads
\begin{displaymath}
     N(r,z,x) = \sum_i 2h q_i J_0\left(\zeta_i \frac{r}{R} \right)
     \exp\left( - \frac{x}{x_i} \right) \Theta(x)
\end{displaymath}
with
\begin{displaymath}
     x_i = \frac{m v \Sigma^0_{\rm ism} L}{2K}
     \frac{\displaystyle \phi \left(\frac{3+a_i}{4}, \frac{3}{2}; 
\frac{V_0L^2}{2K} \right)}
     {\displaystyle \phi\left(\frac{1+a_i}{4}, \frac{1}{2}; 
\frac{V_0L^2}{2K}\right)}
     \;\;\;, \;\;\;
     a_i \equiv \frac{2K}{V_0} \frac{\zeta_i^2}{R^2} + 2
     \; .
\end{displaymath}
where $\phi$ is the confluent hypergeometric function, also denoted 
$_1F_1$ or $M$ in the literature.
The resulting path-length distribution is shown in 
Fig.~\ref{grammages_vent_lin1} for  a linear and a
constant (and discontinuous at $z=0$) wind. It appears that these two situations lead to very 
similar results,
and one can establish a one-to-one correspondence between the parameters $V_c$ 
and $V_0$ of these models (see Fig.~\ref{grammages_vent_lin2}).
It must be noted that the energy losses have not been considered here.
Adiabatic losses are associated to the wind gradient, and
their effect should be different for the two forms of the galactic 
wind: in the constant case, they are confined
to the disk whereas in the linear case, they are present in the whole 
diffusive volume. 


\section{Time-dependent diffusion equation}
\label{sec:time_dependent}

\subsection{Solutions}
\label{subsec:time}

We now turn to the problem of discreteness in time. For that, we must 
solve the time-dependent diffusion problem for an
instantaneous source.
Diffusion occurs independently in the $z$ and $r$ directions.
Neglecting the radial boundary, 
pure diffusion occurs in the radial direction 
and the density can be written as
\begin{displaymath}
     N(r,z,t) = \frac{1}{4\pi K t} e^{-r^2/4Kt}
     \, N(z,t)
\end{displaymath}
where the function $N(z,t)$ satisfies
a time dependent diffusion equation along $z$.
It is convenient to introduce the quantities
$k_{\rm s}=K/hv \sigma n_{\rm ism}$ and $k_{\rm w}=2K/V_c$,
so that $N(z,t)$ is a solution of
\begin{displaymath}
     \frac{\partial N}{\partial (Kt)} =
     \frac{\partial^2 N}{\partial z^2}
     - 2k_{\rm w} \, \mbox{sign}(z) \, \frac{\partial N}{\partial z}
     - 2k_{\rm s} \delta(z) N
     \; .
\end{displaymath}
For point-like and instantaneous sources, the  radial distribution 
in the disk is given by (see Appendix~A)
\begin{equation}
     N(r,z=0,t) = \frac{1}{4\pi Kt} \exp \left(-\frac{r^2}{4Kt} \right)
     \sum_{n=1}^\infty c_n^{-1} e^{-(k_n^2 + k_{\rm w}^2 )  Kt} \sin^2 
        \{ k_n L \}
        \label{sol_disque}
     \; .
\end{equation}
where the discrete set of $k_n$ are the solutions of
\begin{equation}
     k_n \, \mbox{cotan} (k_nL) = - k_{\rm s} - k_{\rm w}
     \label{eq_kn}
\end{equation}
and
\begin{displaymath}
     c_n =  L - \frac{\sin 2k_n L}{2k_n}
     = L + \frac{(\sin k_n L)^2}{k_n^2}
     \left( k_{\rm s} + k_{\rm w} \right)
     \; .
\end{displaymath}
Though it is not immediately apparent in the expression (\ref{sol_disque}), the 
spallations {\em are} taken into account,
through the Eq.~(\ref{eq_kn}) determining the $k_n$.
Moreover, this expression is more general than the form that is 
usually found in the literature, considering
only the effects of escape, which is obtained by replacing the 
right-hand side of Eq.~(\ref{eq_kn}) by 0.

{\color{blue} The result for a source that would accelerate particles for a 
long period of time would be obtained by integrating the above expression 
over the acceleration period. 
Finally, when energy losses are not considered, a whole energy spectrum 
is also simply accounted for by a linear superposition of delta-function
sources.}

\subsection{Interpretation - Relative importance of nearby/recent sources}

The expression (\ref{sol_disque}) can be written as
\begin{displaymath}
    N(r,z=0,t) = \frac{1}{(4\pi Kt)^{3/2}} \exp 
    \left(-\frac{r^2}{4Kt} \right) g(t)
     \; ,
\end{displaymath}
where the function $g(t)$ gives the correction  to the purely diffusive 
case and takes into account all the relevant physical effects.
As such, it depends on the propagation parameters (spallation
cross-section, galactic wind, halo height, diffusion coefficient).
The importance of these effects depends on the position of the source
in the $(r,t)$ plane. 
In particular, the old (large $t$) and remote (large $r$) sources are
more affected by all these effects.
For large values of $t$, $g(t)$ goes rapidly to zero, ensuring the
convergence of the integral over time in Eq.~(\ref{integrale_stat}).
It also makes the function of $r$ obtained by this
integration decreases faster that $1/r$, thus ensuring the
convergence of the integral over $\vec{r}$.

This expression will be applied in a next study to the observed 
distribution of sources.
For now, we want to illustrate the possible effect of discreteness in time.
For that, we divide the sources into several decades in age, and we 
compute their contribution to the total spectrum.
The energy spectrum of a primary species can be obtained by taking into account
the energy dependence of $K$ in expression~(\ref{sol_disque}).
\begin{figure}[h]
     \begin{center}
        \plotone{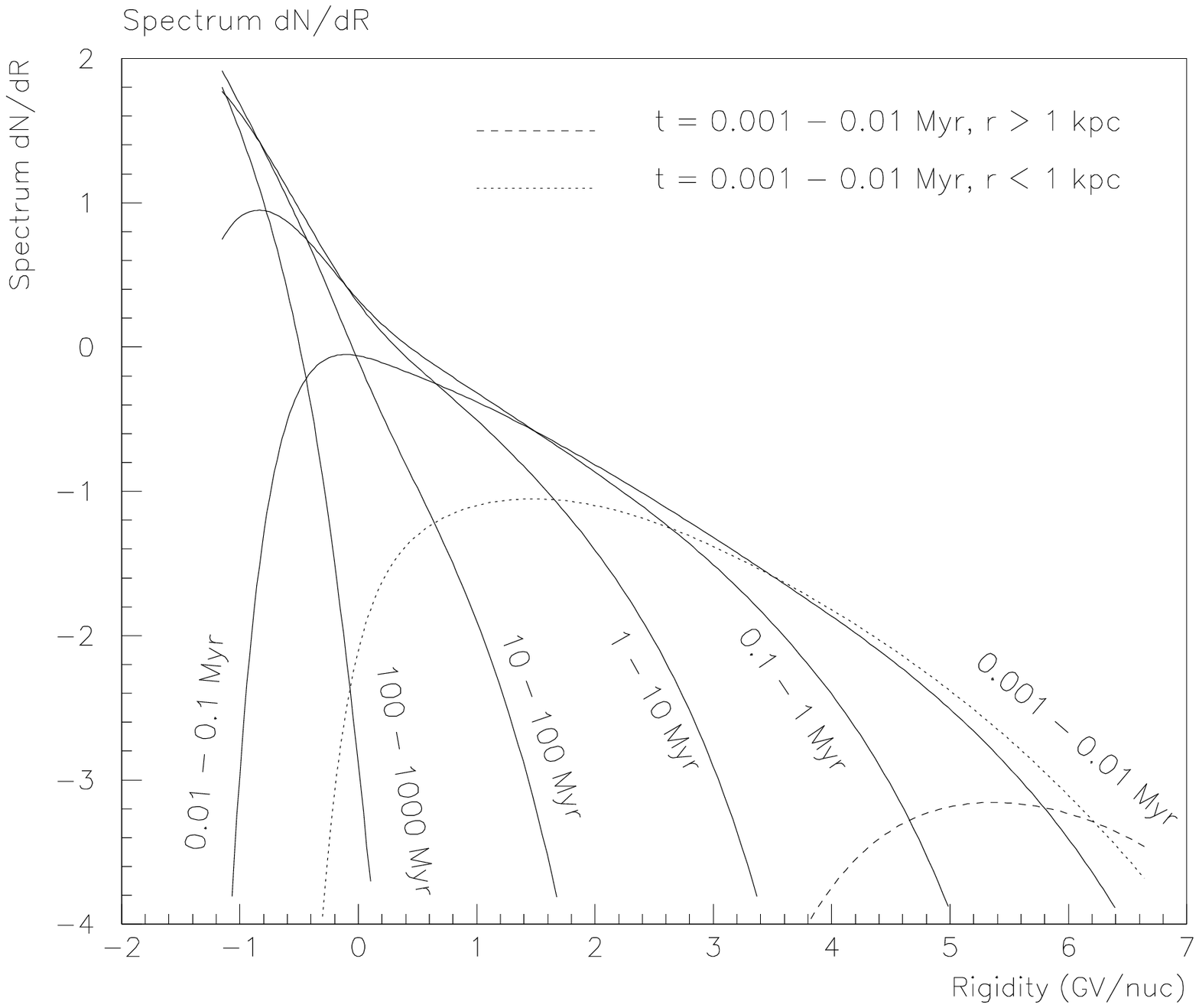}
        \caption{The contribution to the spectrum from different age 
	  classes is displayed for a uniform source distribution.
          In the more recent age class (0.001 Myr $<t<$ 0.01 Myr), 
	  the contribution from nearby sources ($r<1$ kpc) has been 
	  separated from the rest ($r>1$ kpc).
          The effect of time discreteness will be more important in 
	  this recent age class, which contains less sources.
          \label{spectre}}
     \end{center}
\end{figure}
The result is shown in Fig.~\ref{spectre}, where the source distribution 
has been assumed to be uniform in the disk. 
If we denote by $\bar{r}$ the average distance from the sources to 
the Earth, the age $t$ gives the contribution at the energy $E$ for
which $\bar{r} \sim \sqrt{K(E)t}$.
The more recent sources dominate the high energy tail of the spectrum. 
This is where the effect of discreteness in time is expected to be 
the greatest, as the lower decades in age contain the smallest number 
of sources.
For the more recent decade, the sources have been further split into 
nearby ($0.1$ kpc $<r< 1$ kpc) and bulk ($r>1$ kpc). 
For a rate of 3 SN explosions by century in our Galaxy, there should
be about 3 nearby sources in the more recent age decade. 
It is therefore probably important to know the actual position and 
age of these sources, and to correctly model propagation
from these sources, e.g. from Eq.~\ref{sol_disque}.

\subsection{Reformulation of the steady-state model}

The steady-state density results from the continuous superposition of
solutions for instantaneous sources, and thus can be derived from the 
time-dependent solution discussed above:
\begin{displaymath}
     N_{\rm stat}(r,z) = \int_{-\infty}^0 N(r,z,t) \, dt
     \; .
\end{displaymath}
The integration yields
\begin{displaymath}
     N_{\rm stat}(r,z) =   \frac{1}{K} \exp \left(-k_{\rm s} |z| \right)
     \sum_{n=1}^\infty  c_n^{-1} K_0 \left( r\sqrt{k_n^2 + k_{\rm w}^2} \right)
     \sin (k_n L)\,
     \sin \left\{k_n (L-|z|)\right\}
\end{displaymath}
where the Bessel function of the third kind $K_0$ has been introduced.
This expression provides an alternative (but is exactly equivalent) 
to the usual Fourier Bessel expansion over $J_0$ functions.
The functions $K_0$ over which the
development is performed do not oscillate, inducing a faster convergence.
It is thus particularly well suited for sources sharply localized in space,
as point-like sources.
We have checked that this expression is fully equivalent to the
Fourier-Bessel expansion.

\section{Summary and conclusion}

The distribution of cosmic rays sources is not continuous. The granularity
of the distribution has observable effects on the fluxes, spectra and composition, and
thus should be considered when interpreting observed quantities. We have presented
an analytical solution of the diffusion problem for an instantaneous  point source, 
which takes this effect into account when 
the effects of escape through the boundaries $z=\pm L$, convective wind
and spallation are considered. 
The next step is to apply this solution to the observed local distribution of
cosmic ray sources.


\section*{Acknowledgments}
This work has benefited from the support of PICS 1076, CNRS and of
the PNC (Programme National de Cosmologie).



\appendix

\section{Time dependent solution of the diffusion equation}
\label{demo}

We consider diffusion in a cylindrically symmetric box, where both 
disk spallations and galactic wind have been taken into
account. The convection velocity $\vec{V_{c}}$ lies in the vertical
direction and drags the particles outside so that its value is given
by $\mbox{sign}(z) \, V_{c}$ where a constant value for $V_{c}$ has
been assumed. The diffusion equation reads, 
introducing the quantities
$k_{\rm s}=hv \sigma n_{\rm ism}/K$ and $k_{\rm w}=V_c/2K$,
\begin{equation}
{\displaystyle \frac{1}{K} \frac{\partial N}{\partial t}}  = 
\left\{ \frac{1}{r} {\displaystyle \frac{\partial}{\partial r}}
\left( r {\displaystyle \frac{\partial N}{\partial r}} \right)
 +  {\displaystyle \frac{\partial^2 N}{\partial z^2}} \right\}
 - 2k_{\rm w} \, \mbox{sign}(z) \, {\displaystyle \frac{\partial N}{\partial z}}
 - k_{\rm s} \delta(z) N \;\; .
\label{eq_dif_b}
\end{equation}
%
As boundaries in the radial direction $r$ play little
role in our analysis, the galactic disk is modelled as an infinite flat disk 
in the $xOy$ plane. It is furthermore sandwiched by two
confinement layers that extend to $z = \pm L$.
%
The aim of this section is to derive the contribution of a source $S$
located at $\left\{ x = 0 , y = 0 , z = 0 \right\}$ and exploding at
time $t = 0$ to the subsequent cosmic-ray density anywhere else in
the Galaxy at location $P \, \left\{ x , y , z \right\}$. The initial
density reduces to the Dirac distribution
\begin{equation}
N( x,y,z, t=0) \equiv
\delta(x) \, \delta(y) \, \delta(z) \;\; ,
\end{equation}
and we would like to compute it at any time $t > 0$. 
The radial diffusion is independent from diffusion along $z$, and 
is not affected by any of the processes other than diffusion.
As a result, the solution can be factorized into
\begin{equation}
     N ( x,y,z,t) =
     \frac{1}{4 \pi K t} \,
     \exp \left\{ -  \frac{r^{2}}{4 K t} \right\} \,
     n(z,t) \;\; .
\end{equation}
where $n$ is a solution of 
\begin{equation}
     \frac{1}{K} \frac{\partial n}{\partial t} =
     \frac{\partial^{2} n}{\partial z^{2}} -
     k_{\rm w} \, \mbox{sign}(z) \, \frac{\partial n}{\partial z}
     - k_{\rm s}  \delta(z) n \;\; ,
     \label{eq_dif_c}
\end{equation}
The trick is to factorize once again the time $t$ and the vertical
$z$ behaviors so that $n  \equiv f(z) \,  g(t)$, which separates the 
diffusion equation into
\begin{equation}
    g' = - \alpha g
    \;\;\; \mbox{and} \;\;\;
        f''  - 
     2 k_{\rm w} \, \mbox{sign}(z) \, f'
     - k_{\rm s}  \delta(z) f +  \frac{\alpha}{K} f = 0
    \label{separation_t}
\end{equation}
The resulting solution may appear contrived and exceptional. Actually
an infinite set of such functions obtains that turns out to be a natural
basis for the generic solutions to equation~(\ref{eq_dif_c}).
The time behavior amounts to the exponential decrease
$g(t) = \exp \left( - \alpha t \right)$.
The equation on $f$ can be solved for $z>0$ and $z<0$ with the appropriate 
boundary conditions $f(z=\pm L)=0$ as
\begin{displaymath}
    f(z) = A \sinh \left\{k(L-|z|)\right\} \;\; \mbox{when} \;\; V_{c}^{2} - 4 K \alpha > 0
\end{displaymath}
\begin{equation}
    f(z) = A \sin \left\{k(L-|z|)\right\} \;\; \mbox{when} \;\; V_{c}^{2} - 4 K \alpha < 0
        \label{sol}
\end{equation}
with $k^2 = \alpha/K-V_{c}^{2}/4K$. The first possibility does not fulfill the disk crossing
condition (eq.~\ref{eq_implicite_k} with hyperbolic functions).
We therefore disregard it.
We then insert (\ref{sol}) into (\ref{separation_t}). 
Derivation is to be understood in the sense of distributions, because of 
the singularity of $|z|$ in $z=0$. This yields
\begin{displaymath}
    K f''(z) = - \alpha f(z) - 2 K k A \delta(z) \cos kL 
\end{displaymath}
Inserting into (\ref{separation_t}) gives the condition
\begin{equation}
    2k \, \mbox{cotan} (kL) = - k_{\rm s}- k_{\rm w}
    \label{eq_implicite_k}
\end{equation}
The general solution reads
\begin{displaymath}
    N(z,t) = \sum_{n=1}^\infty A_n e^{-\alpha_n t}
    \sin \left\{k_n (L-|z|)\right\}
\end{displaymath}
The functions $\sin \{k_n (L-|z|)\}$ form an orthogonal set, 
and it is found that
\begin{displaymath}
    \int_{-L}^L \sin \left\{k_n (L-|z|)\right\}
    \sin \left\{k_{n'} (L-|z|)\right\}\, dz
    = \delta_{n{n'}} c_n
\end{displaymath}
with
\begin{equation}
    c_n = L - \frac{\sin 2k_n L}{2k_n} 
    \label{def_c_n}
\end{equation}
The $A_n$ are found by imposing that for $t=0$, the distribution is a 
Dirac function, 
\begin{displaymath}
    \delta(z) = \sum_{n=1}^\infty A_n \sin \left\{k_n (L-|z|)\right\}
\end{displaymath}
Multiplying by $\sin \{k_m (L-|z|)\}$ and integrating over $z$ yields
\begin{displaymath}
    A_m =  c_m^{-1} \sin k_m L
\end{displaymath}
so that finally
\begin{equation}
    N(z,t) = \sum_{n=1}^\infty  c_n^{-1} e^{-\alpha_n t} \sin (k_n L)\, 
    \sin \left\{k_n (L-|z|)\right\}
    \label{sol_nonstat_1D}
\end{equation}
and
\begin{equation}
    {\cal N}(r,z,t) = \frac{1}{4\pi Kt} \exp \left(-\frac{r^2}{4Kt} \right)
    \sum_{n=1}^\infty  c_n^{-1} e^{-\alpha_n t} \sin (k_n L)\, 
    \sin \left\{k_n (L-|z|)\right\}
    \label{sol_nonstat_3D}
\end{equation}
The radial distribution in the disk is given by
\begin{equation}
    {\cal N}(r,z=0,t) = \frac{1}{4 \pi Kt} \exp \left(-\frac{r^2}{4Kt} \right)
    \sum_{n=1}^\infty c_n^{-1} e^{-k_n^2 Kt} \sin^2 (k_n L)
    \label{sol_nonstat_3D_disk}
\end{equation}

\section{Path-length distribution for a non homogeneous spallative disk}

This section details the derivation of expression (\ref{sol_non_hom}), giving the grammage distribution in 
the 
case of an arbitrary radial distribution of spallative matter. 
The general method was sketched in \citet{Wallace81} to derive the cosmic ray density profile, and we present 
here a more general version which gives the grammage distribution.

We start from Eq.~(\ref{eq_fond}) for the steady-state case
\begin{displaymath}
     0= K \left\{ \frac{1}{r} \frac{\partial}{\partial r}  \left( r 
{\cal G}(r,z, x) \right)
     + \frac{\partial^2 {\cal G}(r, z, x)}{\partial z^2} \right\}
     + q(r) \delta(z) \delta(x)
     - \frac{\partial {\cal G}(r,z,x)}{\partial x}  v m 
\Sigma^0_{\rm ism} f(r) \delta(z)
\end{displaymath}
with $f(r) \equiv \Sigma_{\rm ism}(r)/\Sigma^0_{\rm ism}$ and 
$\Sigma^0_{\rm ism} = 6.2\, 10^{20}$ cm$^{-2}$ \citep{ferriere98}.
We perform Fourier-Bessel transforms using the $J_0$ functions,
\begin{displaymath}
     {\cal G}(r,z, x) = \sum_{i=0}^\infty {\cal G}_i(z,x) J_0 
\left(\zeta_i \frac{r}{R} \right)
     \;\;\; \mbox{and} \;\;\;
     f(r) {\cal G} (r,z=0, x) = \sum_{i=0}^\infty f_i(x) J_0 
\left(\zeta_i \frac{r}{R} \right)
\end{displaymath}
with
\begin{displaymath}
     {\cal G}_i(z,x) = \frac{2}{J_1^2(\zeta_i)} \int_0^1 \rho J_0 
\left(\zeta_i \rho \right) {\cal G}(\rho R, z, x) d \rho
\end{displaymath}
\begin{displaymath}
     f_i(x) = \frac{2}{J_1^2(\zeta_i)} \int_0^1 \rho J_0 \left(\zeta_i 
\rho \right) f(\rho R)
     {\cal G}(\rho R, z=0,x) = \sum_{j=0}^\infty \alpha_{ij} {\cal G}_j(0,x)
\end{displaymath}
where we have introduced the matrix
\begin{displaymath}
     \alpha_{ij} =   \frac{2}{J_1^2(\zeta_i)}\int_0^1 \rho \,
      J_0(\zeta_i \rho) J_0(\zeta_j \rho) f(\rho r) \, d\rho
\end{displaymath}
These expressions are reminiscent of those of \citet{Wallace81} that 
was dedicated to a perturbative
resolution of the diffusion equation in the presence of an arbitrary 
matter distribution (with no description of the path-length distribution).
The generalized diffusion equation reads
\begin{equation}
     0= K \left\{ \frac{\zeta_i^2}{R^2} {\cal G}_i(z, x)
     + \frac{\partial^2 {\cal G}_i(z, x)}{\partial z^2} \right\}
     + q_i \delta(z) \delta(x)
     - \frac{\partial f_i(x)}{\partial x} v m \Sigma^0_{\rm ism} \delta(z)
     \label{eq_3D_finale}
\end{equation}
The solution for $z\neq 0$ satisfying $f(z = \pm L,x)=0$ is
\begin{displaymath}
     {\cal G}_i(z,x) = {\cal G}_i(0,x) \frac{\sinh \left(\zeta_i 
(L-|z|)/R \right)}{\sinh \left(\zeta_i L/R \right)}
\end{displaymath}
This expression is inserted back in the diffusion equation, taking 
care of the singularity of $|z|$ in 0 which yields
\begin{equation}
     0= {\cal G}_i(0,x)
     + \frac{q_i}{K} \frac{R} {2\zeta_i}\tanh \left( \frac{\zeta_i 
L}{R} \right)  \delta(x)
     - \frac{v m \Sigma^0_{\rm ism} R}{2\zeta_i K} \tanh \left( 
\frac{\zeta_i L}{R} \right)
     \sum_{j=0}^\infty \alpha_{ij} \frac{\partial  {\cal G}_j(0,x)}{\partial x}
     \label{eq_sans_vent}
\end{equation}
The solutions of this linear set of coupled first order differential equations
are
\begin{equation}
     {\cal G}_i(0,x) = \sum_{j=0}^\infty a_{ij} e^{-x/x_j} \, \Theta(x)
     \label{forme_generale}
\end{equation}
where the $x_j$ are the eigenvalues of the matrix
\begin{displaymath}
     A_{ij} = \frac{v m \Sigma^0_{\rm ism} R}{2K\zeta_i}  \tanh 
\left( \frac{\zeta_i L}{R}
     \right) \alpha_{ij}
\end{displaymath}
Indeed, inserting expression (\ref{forme_generale}) in (\ref{eq_3D_finale}),
and using
\begin{displaymath}
     \frac{\partial {\cal G}_j(0,x)}{\partial x} = - \sum_{k=0}^\infty 
\frac{a_{jk}}{x_k}
     e^{-x/x_k} \Theta(x)
     + \delta(x) \sum_{k=0}^\infty a_{jk}
\end{displaymath}
we find an equation that can be separated in a regular part (factor 
of $\Theta(x)$)
and a singular part (factor of $\delta(x)$).
The former reads
\begin{displaymath}
     \sum_{j=0}^\infty a_{ij} e^{-x/x_j}
     - \frac{v m \Sigma^0_{\rm ism} R}{2\zeta_i K} \tanh \left( 
\frac{\zeta_i L}{R} \right)
     \sum_{k=0}^\infty \alpha_{ik}
     \sum_{j=0}^\infty \frac{a_{kj}}{x_j}  e^{-x/x_j}
     = 0
\end{displaymath}
In order for each coefficient of $e^{-x/x_j}$ to be zero, one must have
\begin{equation}
      a_{ij} - \frac{v m \Sigma^0_{\rm ism} R}{2\zeta_i K x_j} 
\tanh \left( \frac{\zeta_i L}{R} \right)
     \sum_{k=0}^\infty \alpha_{ik} a_{kj} = 0
     \label{eq_a_ij_1}
\end{equation}
so that
\begin{equation}
     \det \left[x_j \delta_{ik}- \frac{v m \Sigma^0_{\rm ism} 
R}{2\zeta_i K} \tanh \left( \frac{\zeta_i L}{R} \right)
     \alpha_{ik} \right] = 0
     \label{eigen}
\end{equation}
This shows that the $x_i$ are eigenvalues of $A_{ij}$. This equation 
alone is then not enough to compute the
$a_{ij}$.
An extra relation is provided by the singular part
\begin{displaymath}
     - \frac{v m \Sigma^0_{\rm ism} R}{2\zeta_i K} \tanh \left( 
\frac{\zeta_i L}{R} \right)
     \sum_{j=0}^\infty \alpha_{ij} \sum_{k=0}^\infty a_{jk}
     = -\frac{q_i}{K} \frac{R} {2\zeta_i}\tanh \left( \frac{\zeta_i 
L}{R} \right)
\end{displaymath}
which gives
\begin{equation}
     \sum_{k=0}^\infty  \sum_{j=0}^\infty \alpha_{ij} a_{jk}
     = \frac{q_i}{v m \Sigma^0_{\rm ism}}
     \label{eq_a_ij_2}
\end{equation}
The coefficients $a_{ij}$ are completely set by Eqs~(\ref{eq_a_ij_1}) and 
(\ref{eq_a_ij_2}).

When the galactic wind is taken into account, a more tedious 
derivation shows that Eq.~(\ref{eq_sans_vent}) should be
replaced by
\begin{equation}
     0= {\cal G}_i(0,x)
     + \frac{q_i}{2K}  \left[\frac{1}{r_{\rm w}} + S_i \coth \left( 
\frac{\zeta_i L}{R}
     \right)\right]^{-1}  \delta(x)
     - \frac{v m \Sigma^0_{\rm ism}}{2K} \left[\frac{1}{r_{\rm 
w}} + S_i \coth \left( \frac{\zeta_i L}{R}
     \right)\right]^{-1}
     \sum_{j=0}^\infty \alpha_{ij} \frac{\partial  {\cal G}_j(0,x)}{\partial x}
     \label{eq_avec_vent}
\end{equation}
so that the same results apply, provided that one makes the substitution
\begin{displaymath}
     \frac{R} {2\zeta_i}\tanh \left( \frac{\zeta_i L}{R} \right)
     \rightarrow
     \left(\frac{1}{r_{\rm w}} + S_i \coth \left( \frac{S_i L}{2}
     \right)\right)^{-1}
     \;\;\; \mbox{and} \;\;\;
     S_i = 2\left( \frac{1}{r_{\rm w}^2} + \frac{\zeta_i^2}{R^2}\right)^{1/2}
\end{displaymath}

\section{Linear galactic wind in the steady-state cylindrical disk-halo model}

\paragraph{Resolution of the diffusion equation for a stable primary}

We write $V_c(z)=V_0 z$, and the
steady-state diffusion equation reads (see also \citealp{Bloemen93})
\begin{displaymath}
     0 = \frac{K}{r} \frac{\partial }{\partial r}
     \left\{ r \frac{\partial N}{\partial r} \right\}
     -\frac{\partial N}{\partial z} \left\{
     - K \frac{\partial N}{\partial z} + V_0zN \right\}
     - \sigma \Sigma v \delta(z) N + 2hq \delta(z)
\end{displaymath}
Developing over Bessel functions,
\begin{displaymath}
     -\frac{2h q_i}{K}  \delta(z) = -\frac{\zeta_i^2}{R^2} \, N_i
     + \frac{\partial^2 N_i}{\partial z^2}
     - \frac{V_0z}{K}\,  \frac{\partial N_i}{\partial z}
     - \frac{V_0}{K} N_i - \frac{\sigma \Sigma v }{K} \delta(z) N_i
\end{displaymath}
It is convenient to rewrite this equation in a hermitic differential 
form, to ensure that the solutions form
an orthogonal set of functions (see e.g. \citealt{morse}).
We introduce $n_i = N_i \exp(-k^2z^2/2)$, $\beta = V_0/4K$ and $y=kz$ with $k=\sqrt{V_0/2K}$, 
which yields, in the halo
\begin{displaymath}
     {n_i}'' - n_i \left( a_i + y^2 \right) = 0
\end{displaymath}
where
\begin{displaymath}
     a_i \equiv \frac{2K}{V_0} \frac{\zeta_i^2}{R^2}+ 2
\end{displaymath}
The solutions are of the form, taking into account the condition 
$n_i(z=\pm L)=0$,
\begin{displaymath}
     n_i = B_i \, e^{-k^2z^2/2} \left\{
     \phi\left(\frac{1+a_i}{4}, \frac{1}{2}; k^2z^2 \right) -
     \frac{z}{L}
     \frac{\displaystyle \phi\left(\frac{1+a_i}{4}, \frac{1}{2}; k^2L^2 \right)}
     {\displaystyle \phi \left(\frac{3+a_i}{4}, \frac{3}{2}; k^2L^2 \right)}
     \, \phi \left(\frac{3+a_i}{4}, \frac{3}{2}; k^2z^2 \right)
     \right\}
\end{displaymath}
where $\phi$ is the confluent hypergeometric function, also noted $_1F_1$.
The value of $B_i$ is found by integrating the diffusion equation 
through the disk, so that
\begin{displaymath}
     2 \left. \frac{dn_i}{dz} \right|_{z=0}=
     \frac{\sigma \Sigma^0_{\rm ism} v }{K} n_i(0) - \frac{2h q_i}{K}
\end{displaymath}
This gives $B_i=2h q_i/A_i$ with
\begin{equation}
     A_i = \frac{2K}{L}
     \frac{\displaystyle \phi\left(\frac{1+a_i}{4}, \frac{1}{2}; k^2L^2 \right)}
     {\displaystyle \phi \left(\frac{3+a_i}{4}, \frac{3}{2}; k^2L^2 \right)}
     + v \sigma \Sigma^0_{\rm ism}
\end{equation}
The final solution is thus obtained as
\begin{equation}
     N(r, z) = \sum_i \frac{2hq_i}{A_i} J_0\left(\zeta_i \frac{r}{R} \right)
      \left\{
     \phi\left(\frac{1+a_i}{4}, \frac{1}{2}; k^2z^2 \right) -
     \frac{z}{L}
     \frac{\displaystyle \phi\left(\frac{1+a_i}{4}, \frac{1}{2}; k^2L^2 \right)}
     {\displaystyle \phi \left(\frac{3+a_i}{4}, \frac{3}{2}; k^2L^2 \right)}
     \, \phi \left(\frac{3+a_i}{4}, \frac{3}{2}; k^2z^2 \right)
     \right\}
\end{equation}
The density in the disk is thus given by
\begin{equation}
     N(r, z=0) = \sum_i \frac{2hq_i}{A_i} J_0\left(\zeta_i \frac{r}{R} \right)
\end{equation}
It can be shown that this expression reduces to the usual expressions 
in the case of a vanishing wind.

\paragraph{The path-length distribution}

The dependence in $\sigma$ is very simple and the path-length 
distribution is obtained by inverse Laplace
transform as
\begin{displaymath}
     N(r,z,x) = \sum_i 2h q_i J_0\left(\zeta_i \frac{r}{R} \right)
     \exp\left( - \frac{x}{x_i} \right) \Theta(x)
\end{displaymath}
with
\begin{displaymath}
     x_i = \frac{m v \Sigma L}{2K}
     \frac{\displaystyle \phi \left(\frac{5}{4}+\frac{K}{2V_0} \frac{\zeta_i^2}{R^2}
     , \; \frac{3}{2}; \; \frac{V_0 L^2}{2K} \right)}
     {\displaystyle \phi\left(\frac{3}{4}+\frac{K}{2V_0} \frac{\zeta_i^2}{R^2}, 
     \; \frac{1}{2}; \; \frac{V_0 L^2}{2K} \right)}
\end{displaymath}

\section{Remark about the time evolution of the mean grammage}

The mean grammage of the CR emitted by a single source can be expressed from
the distribution ${\cal G}(\vec{r}, x, t)$ as
\begin{equation}
        \bar{X}(t) \equiv \frac{\displaystyle \int  \!\!\!\int  \!\!\!\int d^3 \vec{r}
        \int_0^\infty dx\, x\, {\cal G}(\vec{r}, x, t)}
        {\displaystyle \int  \!\!\!\int  \!\!\!\int d^3 \vec{r} 
        \int_0^\infty dx\, {\cal G}(\vec{r}, x, t)}
\end{equation}
In this expression, the averaging process is understood to be performed over the whole
spatial distribution of cosmic rays.
The time derivative of this expression yields
\begin{equation}
    \frac{d\bar{X}}{dt} = \frac{\displaystyle \int  \!\!\!\int  \!\!\!\int d^3 \vec{r}
    \int_0^\infty dx\, x\, \partial{\cal G}(\vec{r}, x, t)/\partial t}
    {\displaystyle \int  \!\!\!\int  \!\!\!\int d^3 \vec{r} 
    \int_0^\infty dx\, {\cal G}(\vec{r}, x, t)}
    -\bar{X}\; \frac{\displaystyle \int  \!\!\!\int  \!\!\!\int d^3 \vec{r}
    \int_0^\infty dx\, \partial{\cal G}(\vec{r}, x, t)/\partial t}
    {\displaystyle \int  \!\!\!\int  \!\!\!\int d^3 \vec{r} 
    \int_0^\infty dx\, {\cal G}(\vec{r}, x, t)}
\end{equation}
The denominator is simply $N(t)$, the total number of CR in the diffusive volume.
Using the fact that
\begin{displaymath}
    \frac{\partial {\cal G}}{\partial t} = - \bar{m} \beta c n(\vec{r})
\end{displaymath}
where $n(\vec{r})$ is the interstellar gas density and integrating the first term by parts, we finally find,
\begin{equation}
    \frac{d\bar{X}}{dt} = \bar{m} \beta c \, n_{sn}(t)
    -\bar{X}(t) \, \frac{1}{N}\frac{dN}{dt}
\end{equation}
where $n_{sn}(t)$ is defined as in (\ref{average_higdon}).
The last term is missing in Eq.~4.3 of \citet{Higdon03}. This term is positive and
represents the change in grammage due to the escape of a fraction of cosmic 
rays between times $t$ and $t+dt$. 

As stressed in the text, a quantity which has a greater physical importance to us is
the mean grammage of the CR that reach the Earth at time $t$, i.e.
\begin{equation}
        \bar{x}(\vec{r},t) = \frac{\displaystyle 
        \int_0^\infty dx\, x\, {\cal G}(\vec{r}, x, t)}
        {\displaystyle 
        \int_0^\infty dx\, {\cal G}(\vec{r}, x, t)}
\end{equation}
Following the same procedure as above, its time evolution is given by
\begin{equation}
    \frac{\partial \bar{x}(\vec{r},t)}{\partial t} = \bar{m} \beta c \, n(\vec{r})
    -\bar{x}(\vec{r},t) \, \frac{\partial}{\partial t} \left\{ \ln w(\vec{r},t) \right\}
\end{equation}
where the definition of $w(\vec{r},t)$ is given in (\ref{redef_hidgon}).

\end{document}